\documentclass[12pt]{iopart}

\expandafter\let\csname equation*\endcsname\relax
\expandafter\let\csname endequation*\endcsname\relax

\usepackage{amsmath}
\usepackage{bm}
\usepackage{algorithm}
\usepackage{algpseudocode}
\usepackage{graphicx}
\usepackage{svg}
\usepackage{subcaption}
\usepackage{iopams}
\usepackage{hyperref}
\usepackage{cite}
\usepackage{microtype}

\hypersetup{
colorlinks = true, 	    
linkcolor = magenta,	
citecolor = blue,		
}

\algrenewcommand\algorithmicrequire{\textbf{Input:}}
\algrenewcommand\algorithmicensure{\textbf{Output:}}

\begin{document}

\title[Tuning neural posterior estimation for gravitational waves]{Tuning neural posterior estimation for gravitational wave inference}

\author{Alex Kolmus$^{1}$, Justin Janquart$^{2, 3}$, Tomasz Baka$^{2,3}$,\\ Twan van Laarhoven$^{1}$, Chris Van Den Broeck$^{2, 3}$, Tom Heskes$^{1}$}

\address{$^{1}$ Institute for Computing and Information Sciences (ICIS), Radboud University Nijmegen,
Toernooiveld 212, 6525 EC Nijmegen, The Netherlands}
\address{$^{2}$ Institute for Gravitational and Subatomic Physics (GRASP), Department of Physics,
Utrecht University, Princetonplein 1, 3584 CC Utrecht, The Netherlands}
\address{$^{3}$ Nikhef, Science Park 105, 1098 XG Amsterdam, The Netherlands}

\ead{alex.kolmus@ru.nl}
\vspace{10pt}
\begin{indented}
\item[]\today
\end{indented}

\begin{abstract}
Modern simulation-based inference techniques use neural networks to solve inverse problems efficiently. One notable strategy is neural posterior estimation (NPE), wherein a neural network parameterizes a distribution to approximate the posterior. This approach is particularly advantageous for tackling low-latency or high-volume inverse problems. However, the accuracy of NPE varies significantly within the learned parameter space. This variability is observed even in seemingly straightforward systems like coupled-harmonic oscillators. This paper emphasizes the critical role of prior selection in ensuring the consistency of NPE outcomes. Our findings indicate a clear relationship between NPE performance across the parameter space and the number of similar samples trained on by the model. Thus, the prior should match the sample diversity across the parameter space to promote strong, uniform performance. Furthermore, we introduce a novel procedure, in which amortized and sequential NPE are combined to swiftly refine NPE predictions for individual events. This method substantially improves sample efficiency, on average from nearly 0\% to 10-80\% within ten minutes. Notably, our research demonstrates its real-world applicability by achieving a significant milestone: accurate and swift inference of posterior distributions for low-mass binary black hole (BBH) events with NPE.
\end{abstract}

\section{Introduction}
Inverse problems encompass the challenging task of deducing the underlying causal factors behind observed phenomena in various scientific domains~\cite{akiyama2019first, symes2009seismic, arridge2009optical, baillet2001electromagnetic}. A specific example of such a phenomenon is a gravitational wave (GW) --- coherent, tiny ripples in space-time generated by the acceleration of massive celestial objects such as black holes or neutron stars~\cite{sathyaprakash2009physics}. The observatories of the LIGO-Virgo-KAGRA collaboration~\cite{collaboration2015advanced, acernese2014advanced, akutsu2021overview} regularly observe these GW events~\cite{KAGRA:2021vkt}. The insights derived from analyzing these events have a huge impact on the field of astronomy~\cite{vitale2021first, abbott2019tests, abbott2021population, finke2021cosmology}. To continue progressing, it is crucial to infer the properties of new GW events accurately, and in a timely manner, especially since the computational demands continue to grow as the detectors improve~\cite{abbott2020prospects}. In this introduction, we will give a brief overview of traditional and neural methods for solving inverse problems, focusing on their applicability in GW astronomy.

How does one find the causal factors explaining an observation $x_{obs}$? Traditionally, tackling complex inverse problems involves three components. First, a simulation model is needed to translate event parameters $\theta$ into synthesized observations $x$. Next, a likelihood function $p(x|\theta)$ is determined, and finally, Bayesian inference methods construct a posterior distribution over the parameters $\theta$ given by Bayes' theorem:
\begin{equation}
    p(\theta|x_{obs}) = \frac{p(x_{obs}|\theta) p(\theta)}{p(x_{obs})},
\end{equation}
where $p(\theta)$ is the prior distribution and $p(x_{obs})$ is the evidence. These Bayesian methods often evaluate millions to billions of potential event parameters before converging to the true posterior distribution. Therefore, quick evaluation of the likelihood function is a necessity. However, obtaining such a practical likelihood function $p(x|\theta)$ can be challenging due to mathematical or computational complexity. 

Current GW pipelines built on this traditional framework take a lot of time to run, ranging from hours to a full month depending on the event properties and desired accuracy~\cite{berry2015parameter, Veitch:2014wba, Ashton:2018jfp}.  The primary factor contributing to the runtime is the evaluation of the likelihood, which requires simulating a GW. Simulating a GW can take anywhere from tens of milliseconds to several seconds~\cite{estelles2022time}, dependent on variables like sampling frequency, signal duration, and the chosen simulation algorithm. With the anticipated construction of third-generation detectors~\cite{punturo2010einstein}, alongside the planned upgrades to existing observatories such as LIGO and Virgo~\cite{Kiendrebeogo_2023}, the computational demands are expected to surge. Consequently, the accurate inference of posterior distributions for future GW observations without substantial enhancements poses a growing challenge. As a result, there is a growing interest in alternative methods for GW inference~\cite{zackay2018relative, morisaki2021accelerating, dax2023neural, bhardwaj2023peregrine, lange2018rapid, Gabbard:2019rde, Chua:2019wwt, Williams:2021qyt, Wong:2023lgb, Fairhurst:2023idl, Tiwari:2023mzf, pathak2023fast}.

Simulation-based inference (SBI) methods~\cite{cranmer2020frontier} offer potential alternatives for solving inverse problems in a more computationally efficient manner. These methods approximate the posterior distribution and need only a simulation model. In recent years, neural networks (NNs) have gained considerable prominence in the SBI domain~\cite{lueckmann2021benchmarking}. Due to their expressivity and capacity, NNs can mimic essential components of the Bayesian inference framework: the likelihood~\cite{papamakarios2019sequential}, the likelihood-ratio~\cite{hermans2020likelihood}, and the posterior itself~\cite{rezende2015variational}. The neural likelihood ratio and neural posterior methods can be trained either for a single event or for any possible event from the prior distribution; these modes are respectively referred to as non-amortized and amortized inference. The latter takes longer to train and is potentially less accurate but the computational burden is paid in advance and only once. Consequently, amortized inference is preferred when faced with low-latency or high-volume challenges. Of special interest is amortized neural posterior estimation (NPE)~\cite{rezende2015variational}, where one trains a conditional neural density estimator to transform a simple, well-understood distribution into an approximate posterior $Q(\theta|x)$. To our knowledge, this is the only neural SBI method that does not require any subsequent Bayesian or variational inference steps to construct an approximate posterior and thus allows for sub-second inference~\cite{green2021complete}. 

In NPE, an NN parameterizes an approximate posterior distribution over the event parameters. The NN is trained by feeding it simulated observations $x$ and iteratively increasing the likelihood of the true parameters $\theta$ in the predicted distribution. While mixture density networks~\cite{bishop1994mixture} and normalizing flow (NF) models ~\cite{rezende2015variational} are both commonly used in NPE, our focus in this work is exclusively on NF models due to their high expressivity. NFs consist of a sequence of differentiable, bijective functions with parameters defined by NNs. These functions can transform a simple distribution into a complex one while accurately tracking the likelihood via the change of variables theorem. The loss function commonly used in NPE is the forward KL divergence, which is equivalent to maximum likelihood estimation for NFs~\cite{papamakarios2021normalizing}. Due to the mode-covering property of the forward Kullback-Leibler (KL) divergence, the approximate posterior should always cover the true posterior\cite{minka2005divergence}. This property enables using importance sampling to converge to the true posterior when a known likelihood function is available. As we will see, NPE followed by importance sampling produces similar results as traditional methods for significantly reduced computational costs~\cite{midgley2022flow, dax2023neural}.

While NPE holds promise as an alternative for full GW inference, certain challenges need to be addressed. First, NPE can struggle with generalizing across the entire parameter space. As we shall demonstrate in section~\ref{sec:effective_priors}, even for simple problems, NPE can have poor sample efficiency for specific subsets of the data. We hypothesize and characterize a correlation between the performance of an NPE model for a specific event and the number of similar samples it has been trained on. The reasoning behind this hypothesis is that NNs learn from examples. Effective training thus demands a prior that exposes the NN to diverse samples, which often does not correspond to the uninformative prior, but an effective one. Second, NPE models struggle to be competitive with Bayesian inference when they need to learn large numbers of high-dimensional observations, producing posteriors that appear correct but are wider than their Bayesian counterparts. Extended training can compensate to some extent, but does not scale well. In section \ref{sec:fine_tuning} we propose fine-tuning of trained NPE models for single instances of the problem. This procedure optimizes the NPE model by self-sampling and correcting these samples with an importance-weighted loss function. By switching from learning all possible events to only a single instance, the problem becomes a lot easier to optimize for. To demonstrate the improvements offered by switching to effective priors and fine-tuning, section \ref{sec:gravitational} shows that we can infer previously inaccessible low-mass parameter ranges\footnote{Down to a chirp mass of 5 solar masses.} for binary-black hole (BBH) mergers observed with GWs. To our knowledge, the inference of posterior distributions for low-mass BBH events remains beyond the reach of existing SBI algorithms~\cite{dax2023neural, delaunoy2020lightning, bhardwaj2023peregrine}. 

\clearpage
\section{\label{sec:exp_section}Experimental setup}
To investigate the behavior of NPE, we first start with a simple toy problem. This section will first describe our toy problem: coupled-harmonic oscillators, an ideal toy problem for three reasons: (1) they are computationally inexpensive to generate, (2) there is a known and cheap likelihood function, which makes importance sampling straightforward, and (3) like gravitational wave observations, they are a time series and have correlated channels. This section will end with a description of our NPE model and training setup.

\subsection{\label{sub:description}Toy problem description}
\begin{figure}[h]
    \centering
    \includegraphics[width=\linewidth]{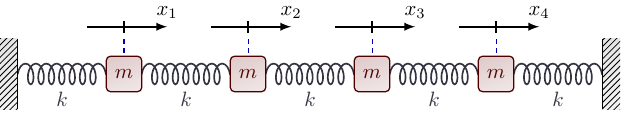}
    \caption{An illustration of the coupled-harmonic oscillators used in the toy problem. There are four oscillators, each has a mass $m$, and they are connected via springs with spring constant $k$. Their displacement from rest position $x_u(t)$ is measured along the horizontal plane.}
    \label{fig:coupled-oscillator}
\end{figure}

We study a linear chain containing four oscillators moving along a single axis, as illustrated in Figure~\ref{fig:coupled-oscillator}. Each oscillator has a mass $m$ and is connected to its neighbors by springs with a spring constant $k$. The first and last oscillators in the chain are attached to rigid walls by springs on their left and right sides. The system's dynamics can be expressed in terms of normal modes $v$. This expression derived in~\cite{torre2012foundations} where the displacement over time $x_u(t)$ of oscillator $u$ is expressed as a sum over the four normal modes with known amplitudes $a_v$ and phases $\phi_v$:
\begin{equation}
    x_u(t) = \sum_{v=1}^4 |a_v| \sin \, \left( \frac{v}{5} u \pi \right) \cos \, \left(2\sqrt{\frac{k}{m}}  |\sin \left( \frac{v\pi}{10} \right)| t + \phi_v \right)
\end{equation}
Conversely, given the displacements over time, the amplitudes and phases of the normal modes can be determined. The described toy problem has a 10-dimensional parameter space: mass $m$, spring constant $k$, and for each normal mode an amplitude $a_v$ and phase $\phi_v$. 

To introduce uncertainty into the inverse problem, we incorporate white noise into the observed displacements $x_{obs}(t)$, and discretize these into $\bm{x}_{obs}$\footnote{Bold symbols indicate vectors.} depending on the sampling frequency. This allows us to use the simplest version of the Whittle likelihood function~\cite{whittle1953estimation}, given by:
\begin{equation}
    p(\bm{x}_{obs}|\bm{\theta}) \propto \exp\left(-\sum_{t, u} \frac{(\bm{x}_{obs, u} - \bm{s}_u(\bm{\theta}))^2}{2}\right),
\end{equation}
where $\bm{\theta}$ represents the inferred mass, spring constant, amplitudes, and phases, and $\bm{s}_u(\bm{\theta})$ denotes a clean simulated signal parameterized by $\bm{\theta}$. This likelihood function measures how well the residuals, the observation minus the simulation, match a standard normal distribution. A set of $\{\bm{\theta}_i\}$ for which the residuals resemble white noise should explain the observation well. To evaluate the performance of an NPE model, ideally, we would calculate the KL divergence between the exact Bayesian posterior and the posterior predicted by the NPE model. However, the computational costs would be excessively high for all the experiments in this paper. Instead, the sample efficiency $\eta$ is used to quantify the performance of the NPE model. For $n$ drawn samples from the NPE model, the sample efficiency is defined as:
\begin{equation}
\eta = \frac{\left(\sum_{i=0}^n w_i\right)^2}{n \sum_{i=0}^n w_i^2} = \frac{n_{\text{eff}}}{n},
\end{equation}
where $w_i$ represents the ratio between the Whittle likelihood and the likelihood given by the NPE model for the $i$th sample. And $n_{\text{eff}}$ is the Kish effective sample size~\cite{kish1965survey}. If $n$ is sufficiently large and the support of our approximate distribution covers the support of the true distribution, we can interpret the sample efficiency as a quality measure of the approximate distribution. NF models trained with the forward KL-divergence are in general mode-covering and can generate thousands of posterior samples within a second, satisfying these requirements. 

\subsection{\label{sub:model}NPE model specification and training}
\begin{figure}
    \centering
    \includegraphics[width=0.8\linewidth]{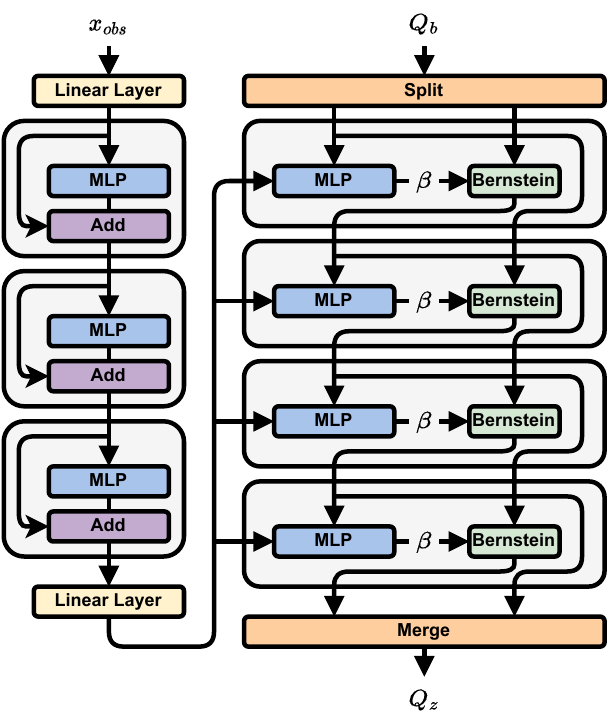}
    \caption{A schematic of our NPE model. The flow through the schematic is made explicit by arrows. The left side of the schematic shows the context network, which is a residual network consisting of three residual blocks. The right side shows the NF model which transforms a simple distribution $Q_b$ into an approximate posterior distribution $Q_z$. The NF model consists of four coupling layers, each conditioned by the output of the context network. The method of conditioning is discussed in more detail in~\ref{app:fine_tuning_conditioning}. Each coupling layer has two inputs, a dynamic half and a static half. The static half is used as input into an MLP which produces the $\beta$ coefficients for the Bernstein polynomial, which transforms the dynamic half. At the end of the coupling layer the dynamic and static halves trade places for the consecutive coupling layer.}
    \label{fig:npe_model}
\end{figure}

As can be seen in Figure~\ref{fig:npe_model}, the NPE model is a combination of two models:

\textbf{(1) The context model} transforms the time series into a neural representation. It begins with a linear transformation, followed by three residual blocks, and ends with another linear transformation to produce the neural representation. Although it has a consistent structure across experiments, the dimensions of the linear layers can change to accommodate longer or more complex time series. The specific dimensions of the context network for each experiment can be found in~\ref{app:experimental_details}.

\textbf{(2) The NF model} transforms a base distribution $Q_b$ into a complicated distribution $Q_z$. The NF model builds this transformation via a series of coupling layers~\cite{dinh2015nice}. A coupling layer with index $l$ divides the input into two halves: a dynamic $\bm{b}^l_{i}$ and static $\bm{b}^l_{j}$, where the static half acts as a condition for the transformation of the dynamic half. The transformation is a function $f$, which has to be differentiable and bijective in its first parameter. It is typically a monotonically increasing polynomial whose coefficients $\beta$ are generated by an NN $g$ with parameters $\bm{\tau}_l$. The input to $g$ is the static half, and possibly a context vector $c$. The output of such a coupling layer is
\begin{align}
    \bm{b}^{l+1}_{i} &= f(\bm{b}^{l}_{i}, g(\bm{b}^{l}_{j}; \bm{\tau}_l)) = f(\bm{b}^{l}_{i}, \beta^l) \\
    \bm{b}^{l+1}_{j} &= \bm{b}^{l}_{j}.
\end{align}
By alternating which dimensions are dynamic and which are static in consecutive coupling layers, the model can represent a flexible distribution over the parameters. The entire series of coupling layers is denoted $S$ and the corresponding set of parameters is denoted $\bm{\psi}$. The principle of a normalizing flow is based on the equivalence relation between $Q_z$ and $Q_b$ via the change of variables theorem:
\begin{equation}
    Q_z(\bm{z}| \bm{\psi}) = Q_b(\bm{b}) \, | \, \det \, \mathbf{J}_S(\bm{b}; \bm{\psi}) \,|^{-1} \text{ where } \bm{z} = S(\bm{b}).
\end{equation}
Here, $\mathbf{J}_S(\bm{b}; \bm{\psi})$ represents the Jacobian of the transformation function. One can optimize $Q_z$ to approximate an (unnormalized) target distribution $P(\bm{z})$ with $Q_z(z|\bm{\psi})$ by minimizing the forward KL-divergence $D_{KL}(P(\bm{z})|Q_z(\bm{z}|\bm{\psi}))$, which for NF models is equivalent to fitting $Q_z(\bm{z}|\bm{\psi})$ by maximum likelihood estimation~\cite{papamakarios2021normalizing}. The loss function for a single sample reads:
\begin{equation}
    L(\bm{z}| \bm{\psi}) = -\log(Q_z(\bm{z}| \bm{\psi})).
\end{equation}

The base distribution of our NF model is a truncated standard normal distribution. We went with a truncated distribution since they match naturally with the boundaries of the parameter space, for example, the phase is bounded between 0 and 2$\pi$. For our transformation function, we choose Bernstein polynomials~\cite{ramasinghe2021robust}, which are both highly expressive and robust, regardless of noise or polynomial order. These qualities allow us to build a relatively shallow, yet highly expressive NF model. It is also faster to train and has a smaller memory footprint, compared to the conventional RQ-spline NF models~\cite{durkan2019neural}. As we shall see in section~\ref{sec:fine_tuning}, a fast NF model is very beneficial if low latency is desired. For all of the experiments, the Bernstein polynomials are parameterized by a shallow multi-layer perceptron (MLP). The details of the NF model for each experiment can be found in~\ref{app:experimental_details}.

\clearpage
\section{\label{sec:effective_priors}Effective priors for NPE}
The relationship between the size of the training dataset and NN performance remains a topic of ongoing research~\cite{zhang2021understanding, kaplan2020scaling, power2022grokking}. However, the general sentiment is that increasing the size of the dataset improves performance. Conversely, NNs do not perform well at inference time for input which it has not been sufficiently trained on. Neural simulation-based inference relies on training the model with simulated data originating from a chosen prior. Conventionally, one uses an uninformative prior to mirror the Bayesian inference framework. In this section, we argue that to train a robust and accurate NPE model one has to choose the prior such that the model trains on an as diverse set of samples as possible. In other words, the prior should be effective in training the NN.

We used the toy problem for all the experiments in this section. We simulated observations from the coupled-harmonic oscillator of two seconds at a sampling frequency of 128 Hz. The simulated time series is a mix of four sinusoids, whose frequencies $f_v$ are proportional to $\sqrt{{k}/{m}}$. A change in mass does not translate into a linear response in frequency. It implies that with a uniform prior on $m$, there is more data with low frequencies. Ideally, for any sample drawn from the prior, the number of similar samples is roughly equal.

\begin{figure}[ht]
    \centering
    \includegraphics[width=0.49\linewidth]{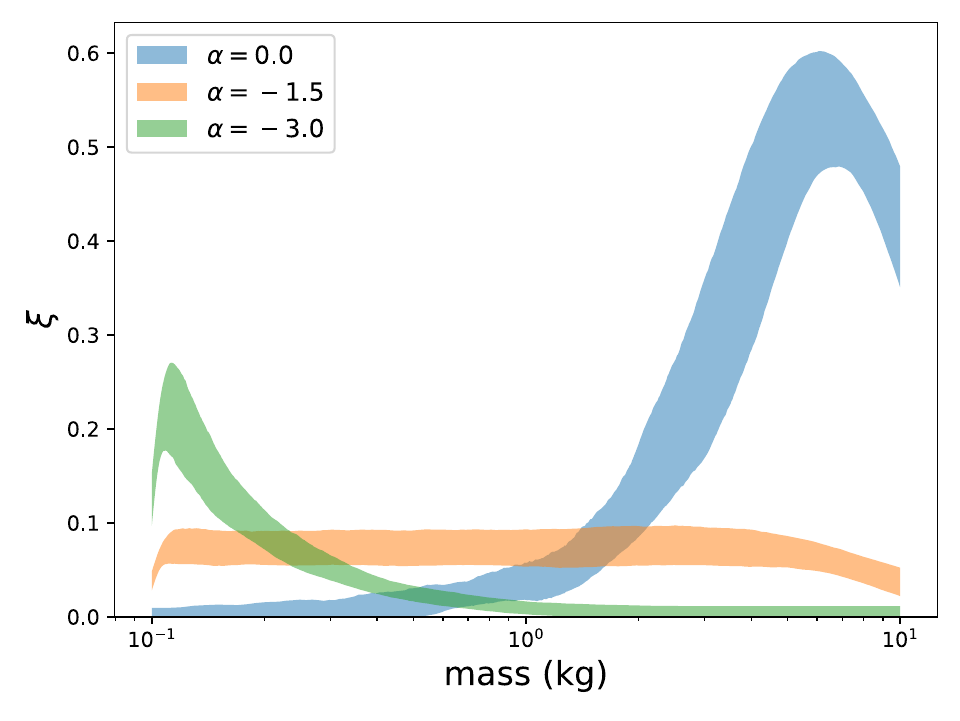}
    \hfill
    \includegraphics[width=0.49\linewidth]{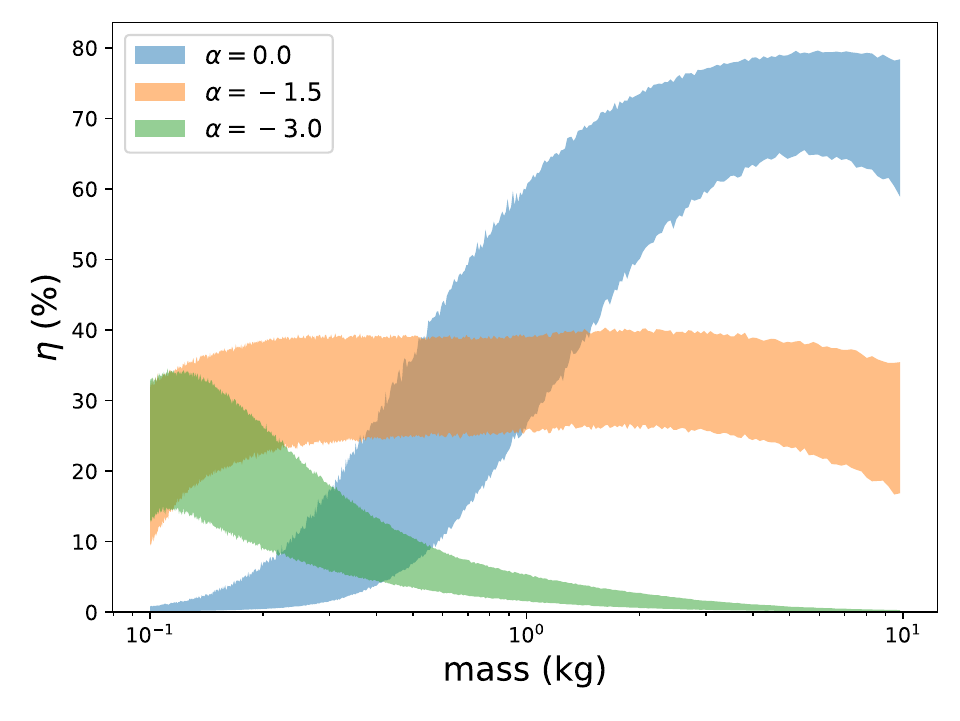}
    \caption{A comparison between the sample exposure for different priors and the sample efficiency of the corresponding NPE models. \textbf{Left.} The sample exposure at a specific mass for three different priors. The priors consist of a uniform prior (blue), a power law with an exponent of -1.5 (orange), and a power law with an exponent of -3.0. To cover the influence of the other parameters, we compute the sample exposure across the mass with 1000 different instances of $\bm{\theta}_{rest}$. The band shows the central 50\% of computed sample exposures. \textbf{Right.} The sample efficiency for three NPE models trained with the three different priors, the shown band covers the central 50\%. Although the sample exposure and sample efficiency do not match exactly, there is a clear correspondence between them.}
    \label{fig:sample_cohesion}
\end{figure}

To quantify the similarity between time series we use the cosine similarity, also known as the match in GW astronomy. By keeping one sample as a constant argument and drawing the other from a chosen prior, we estimate the NN's exposure to the reference sample. Here, the analysis is limited to mass because the similarity between samples changes the most across this dimension, and is, therefore, the most troublesome to infer correctly. To determine the number of similar signals that the model sees as a function of the prior $p(m)$ we define the sample exposure as
\begin{equation}
     \xi(m, \bm{\theta}_{rest}) =  \mathbb{E}_{m'\sim p} \left[ \frac{\bm{s}(m, \bm{\theta}_{rest})^T \bm{s}(m', \bm{\theta}_{rest}) }{||\bm{s}(m, \bm{\theta}_{rest})|| \, ||\bm{s}(m', \bm{\theta}_{rest})|| } \right],
\end{equation}
where \textbf{$\bm{\theta}_{rest}$} are the all parameters except the mass $m$. We approximated $\xi (m, \bm{\theta}_{rest})$ by sampling 1000 equidistant points from the inverse cumulative density function of $p(m)$. On the left side of Figure~\ref{fig:sample_cohesion}, we show the result of this calculation for three different priors: a power law\footnote{A power law distribution with power $\alpha$ is defined as $p_{\alpha}(x) = x^{\alpha} / A$ where $A$ is a normalization constant.} with $\alpha=-3$, $\alpha=-1.5$, $\alpha=0$. The right side shows the sample efficiency of the corresponding NPE models. The sample exposure seems to align well with the sample efficiency. It seems that NPE models demonstrate strong performance only when they have been exposed to a sufficient number of (similar) observations. If one desires a stable performance over the entire parameter space, it is critical that one chooses a prior that gives uniform sample exposure.

\begin{figure}[hb]
    \centering
    \includegraphics[width=0.49\linewidth]{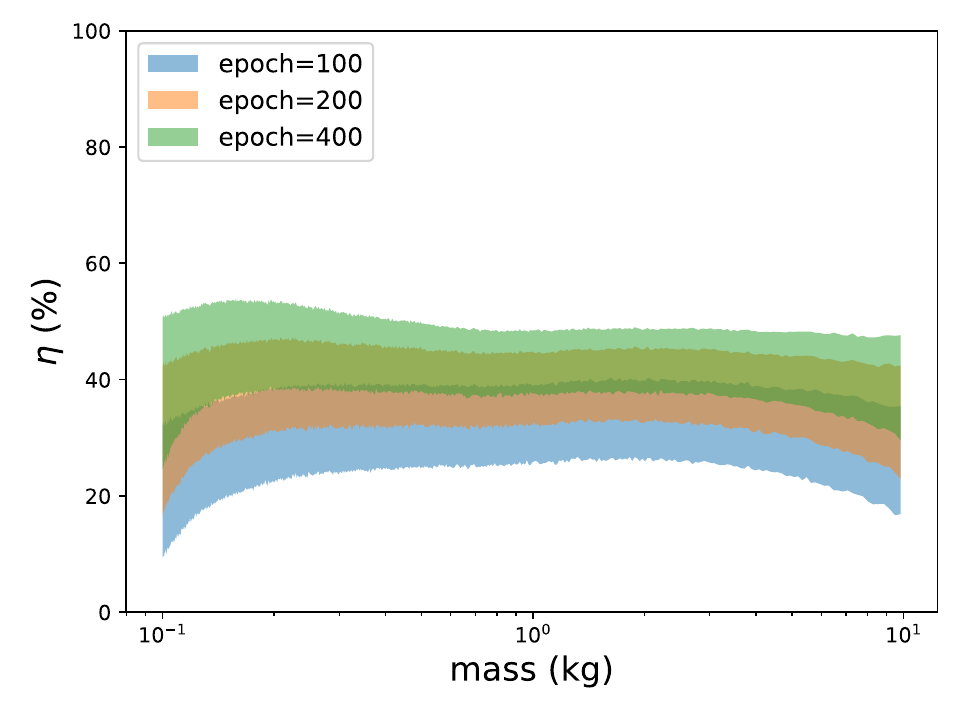}
    \hfill
    \includegraphics[width=0.49\linewidth]{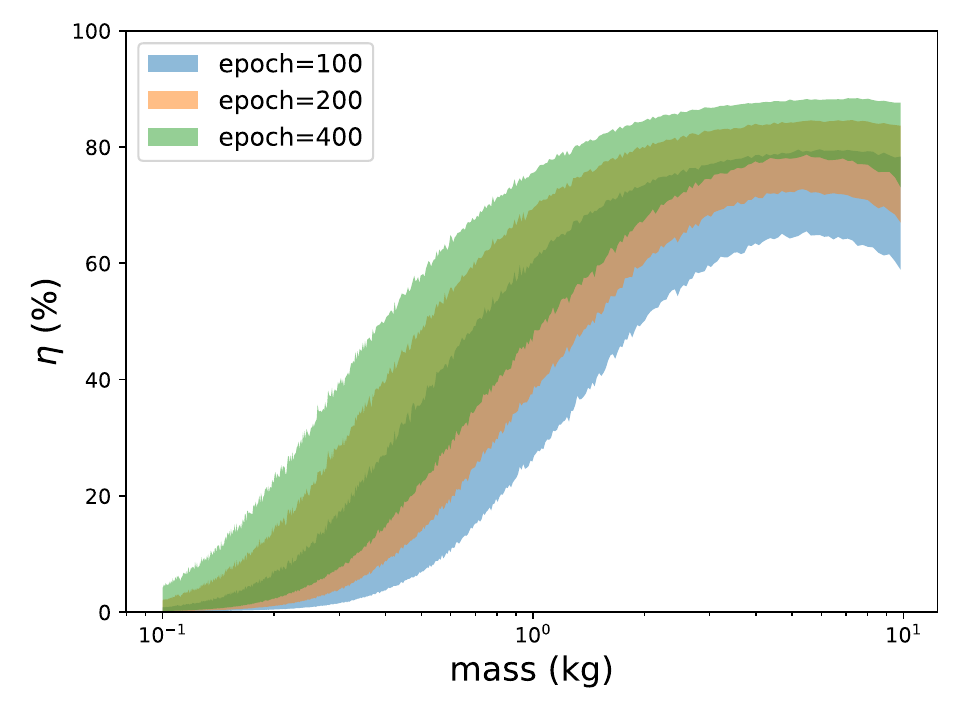}
    \caption{The sample efficiency for different training durations: 100 (blue), 200 (orange), and 400 (green) epochs. The shown bands cover the central 50\%. \textbf{Left.} An NPE model trained with power law distribution ($\alpha = -1.5$) as mass prior. \textbf{Right.} An NPE model trained with uniform distribution as mass prior. Training for longer improves sample efficiency regardless of prior. Despite the improvements in sample efficiency, longer training does not give satisfactory performance for small masses when using a uniform prior. Thousands of epochs are probably needed to guarantee sufficient sample efficiency across the mass range.  Only the power law distribution shows stable performance over the entire mass range regardless of training iteration. }
    \label{fig:unif_vs_power}
\end{figure}

To further validate the hypothesis that the model's performance improves with increased exposure to similar samples, we conducted a second experiment. Two NPE models, one with a uniform mass prior and the other with a power law prior of $\alpha=-1.5$, were trained for 100, 200, and 400 epochs\footnote{A single epoch is 5000 updates with a batch of 1024.}. Figure~\ref{fig:unif_vs_power} presents the sampling efficiencies obtained from this extended training. The results demonstrate a significant enhancement in performance for both $\alpha=-1.5$ and $\alpha=0$ as the training duration increases. This finding further supports the notion that the NPE model functions optimally when it has encountered a sufficient number of similar observations. However, it also indicates that training for longer has diminishing returns. In the next section, we propose a scheme to overcome this problem.

\clearpage
\section{\label{sec:fine_tuning}Fine-tuning neural posterior estimation}
As demonstrated in the previous section, exposure to a diverse set of samples is necessary to ensure a strong NPE model. However, when dealing with an enormous parameter space, obtaining adequate exposure can require billions of samples. To accurately store the massive volume of information, the NPE model must become bigger and thus will be slower to train, requiring more sophisticated hardware. At a certain point, NPE will no longer be viable due to the training requirements. In this section, we address the complications arising from large parameter spaces and present a fine-tuning procedure designed to maintain high sample efficiency, regardless of the parameter space's scale.

\subsection{\label{sub:large_inference_spaces}Challenges in large parameter spaces}
To illustrate the challenges posed by larger parameter spaces, we repeat the experiment of the previous section with a longer duration signal. The duration was changed from two to twenty seconds. This adjustment decreased the sample exposure by a factor of ten. To counteract the decrease in exposure, the NPE model needs to be trained ten times longer. Moreover, the NPE model was given a context network that was a factor ten wider than the original one. Despite these changes, the sample efficiency was only 0.02\% after 100 epochs, and 0.07\% after 1000 epochs. This is significantly worse than the 20--30\% sample efficiency achieved in Section~\ref{sec:effective_priors}. 

While the NPE model was still able to approximate the posterior distribution for the extended twenty-second signal (see Figure~\ref{fig:oscillator_corner}) its predictions were significantly wider than its importance-sampled counterpart. This widening suggests that the NPE model was not able to extract all the information from the signal, despite being trained on roughly 5 billion unique samples. Naturally, a signal with a longer duration contains more information, and therefore a tighter posterior distribution. This is reflected in the decreased sample exposure, but cannot account for the significant drop in performance. We attribute the lower performance to the inherent difficulty of accurately storing more and much higher-dimensional time series. Traditional methods can still find the correct posterior by running for longer. Altering amortized NPE such that iterative improvements post-prediction are possible might be the solution for large parameter space problems.

\subsection{\label{sub:fine_tuning}Fine-tuning procedure}
As is evident from the last subsection, learning the posterior distributions for all possible events becomes increasingly harder as the parameter space grows or the sample exposure decreases. To circumvent these difficulties we propose switching back to a non-amortized setting after training the NPE model. From now on, we will refer to the optimization of a trained, amortized NPE model for a single observation $\bm{x}_{obs}$ as fine-tuning. Fine-tuning makes learning the posterior distribution more manageable for two reasons. First, the NPE model only needs to train on parameters that produce simulations resembling $\bm{x}_{obs}$. These can be sampled from the amortized model. Second, the NF only needs to learn a single posterior distribution it already roughly approximates. In summary, fine-tuning enables the NPE model to quickly learn the posterior distribution by being more sample-efficient and simplifying the objective.

We will now discuss the implementation of the fine-tuning, outlined in Algorithm~\ref{algo:fine_tuning}. To switch from an amortized setting to a non-amortized setting, the context vector $\bm{c}$ is calculated by passing observation $\bm{x}_{obs}$ through the context model and using it as a static condition for the NF model. For clarity, we define a new NPE model $R(\bm{\theta}|\bm{\psi}', \bm{c})$ whose parameters $\bm{\psi}'$ are initialized with the parameters $\bm{\psi}$ of $Q(\bm{c})$. The remainder of the fine-tuning procedure operates in three steps:
\begin{enumerate}
    \item Generate samples $\bm{\theta}_i$ from distribution $R(\bm{\theta}|\bm{\psi}', \bm{c})$ and calculate the true posterior probability $p(\bm{\theta}_i|\bm{x}_{obs})$ by multiplying the Whittle likelihood $p(\bm{x}_{obs}|\bm{\theta}_i)$ and the prior $p(\bm{\theta}_i)$.
    \item Calculate the posterior probability of $\bm{\theta}_i$ under the NF model $R(\bm{\theta}_i|\bm{\psi}', \bm{c})$.
    \item Update $\bm{\psi}'$ with the $\chi^2$-divergence as loss function:
    \begin{equation}
        L(\bm{\theta}_i; \bm{x}_{obs}, \bm{\psi}') = - \left(\frac{p(\bm{\theta}_i|\bm{x}_{obs})}{R(\bm{\theta}_i|\bm{\psi}', \bm{c})}\right)^2  \log(R(\bm{\theta}_i| \bm{\psi}', \bm{c}))
    \end{equation}
\end{enumerate}
The loss function, as introduced in reference~\cite{muller2019neural}, uses the square of the importance weight rather than the regular importance weight. This approach serves to minimize the variance of importance weights and discourages the importance weights from becoming too big, leading to improved convergence and sample efficiency. During fine-tuning most of the time is consumed by running simulations to calculate the Whittle likelihood. By repeating steps (ii) and (iii) for the same samples generated in step (i), we can cut down on simulation time and still improve our model. In our experiments, we could repeat steps (ii) and (iii) at least ten times while still having a similar loss progression as without any repeated steps. We will refer to completing steps (i), (ii), and (iii) --- including repetitions --- as a cycle. For harder problems, more cycles, and samples, are needed to reliably converge to the correct posterior. Increasing the number of samples generated in step (i) has been sufficient to always find the correct posterior distribution, regardless of multi-modality or the quality of the initial posterior prediction. However, this does increase the time needed to fine-tune the model. As we will see in section~\ref{sec:gravitational} we can mitigate most issues with multimodalities by redefining $R$, saving a lot of time.

Fine-tuning has a close resemblance to sequential NPE methods~\cite{papamakarios2016fast, greenberg2019automatic, lueckmann2017flexible, deistler2022truncated}. Both use self-sampling to generate samples and a (pseudo-)importance weight to update the model. However, sequential NPE models seem to shun the use of amortized models as initial priors and use their own likelihood estimates as a replacement for the true likelihood. The importance ratio is then calculated between sequential iterations of the model, potentially requiring many rounds to converge. Moreover, without using an amortized model, the initial sample quality can be poor, potentially missing part of the posterior due to strong non-convex likelihood landscapes. Or requiring long run-times to explore the parameter space. All of these issues are mitigated by using fine-tuning. To our knowledge, this is the first time amortized and non-amortized SBI have been combined.

\begin{figure}[htb]
    \centering
    \includegraphics[width=\linewidth]{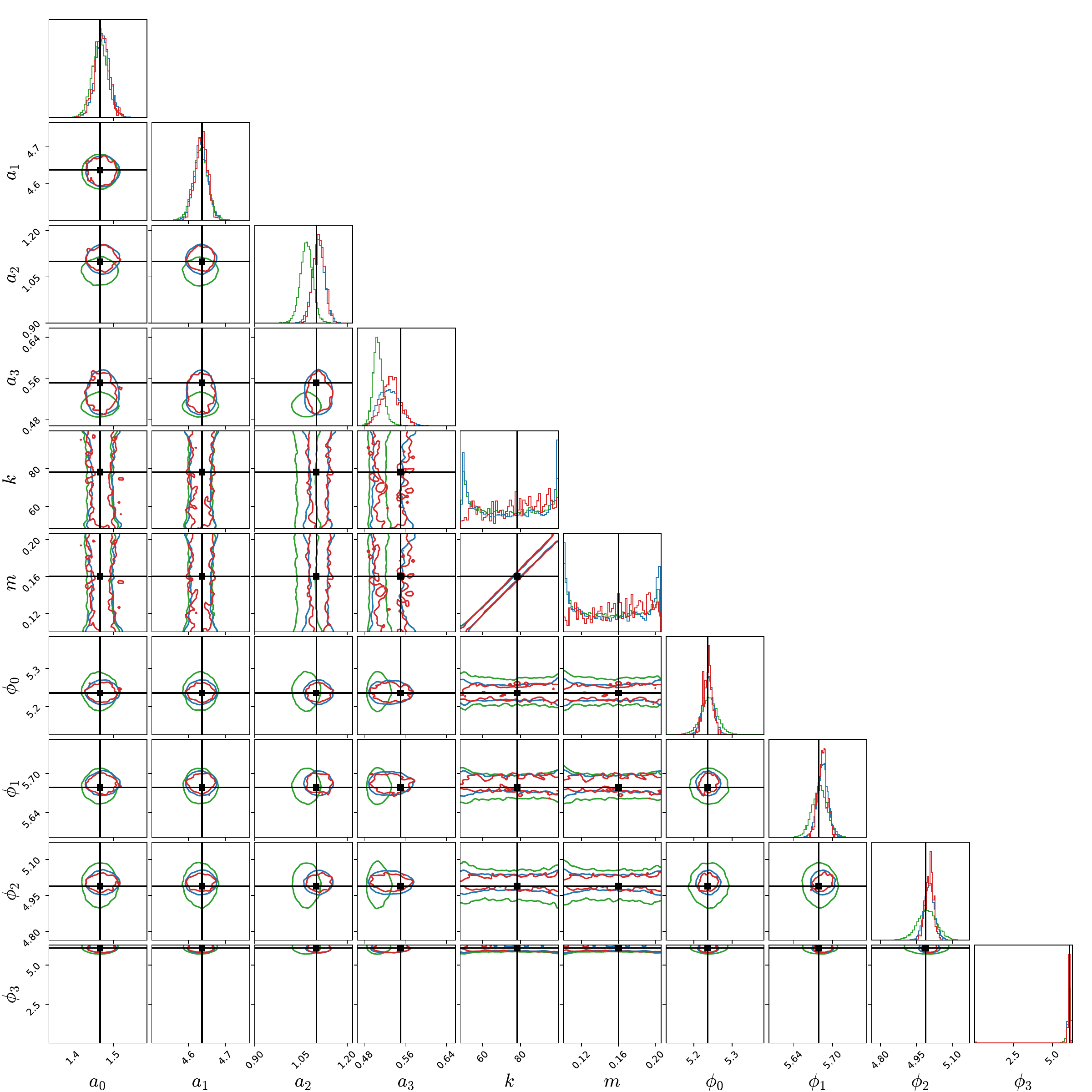}
    \caption{The predicted posterior distribution for 20 seconds long signal, shown as
1D histograms and 2D contour plots. There are three posterior predictions, the original NPE model (green), the fine-tuned NPE model (blue), and the importance-sampled posterior (red). The contours represent the 90\% confidence area. The sample efficiency of the green posterior is 0.1\%, and of the blue posterior it is 3.5\%. The improvement was achieved in five seconds.}
    \label{fig:oscillator_corner}
\end{figure}

The results of our fine-tuning procedure, depicted in Figure ~\ref{fig:oscillator_corner}, underline its ability to improve the sample efficiency of NPE models. The green area represents the posterior predicted by the original NPE model, while the blue area represents those predicted by the fine-tuned NPE model, and the red area depicts the true posterior, derived through importance sampling of the fine-tuned distribution. The fine-tuning was performed for 10 cycles, with a batch size of 10240, 10 repetitions, and finished within five seconds. Fine-tuning brought the sample efficiency from 0.1\% to 3.5\%, sufficient to extract the true posterior with importance sampling. To reach higher sample efficiencies we need to add more NF layers to the model, see~\ref{app:fine_tuning_conditioning}. In the next section, we will use the fine-tuning procedure to quickly find an accurate posterior for GW events.

\begin{algorithm}
    \caption{Fine-tune NPE model}\label{algo:fine_tuning}
    \begin{algorithmic}
        \Require observation $\bm{x}_{obs}$, context model $M$, parameters of pre-trained NPE model $\bm{\psi}$
        \State $\bm{c} \gets M(\bm{x}_{obs})$ \Comment{Generate the context vector}
        \State $\bm{\psi}' \gets \bm{\psi}$
        \For{i in 1..cycles}
            \State $\bm{\theta}_i \sim R(\bm{\theta} |\bm{\psi}', \bm{c})$
            \State $p_i \gets p(\bm{x}_{obs}|\bm{\theta}_i) \, p(\bm{\theta}_i)$
            \For{j in 1..10}
                \State $w_i \gets p_i / R(\bm{\theta}_i|\bm{\psi}', \bm{c})$ \Comment{No gradients are calculated}
                \State $L \gets -w_i^2 \log(R(\bm{\theta}_i|\bm{\psi}', \bm{c}))$
                \State $\bm{\psi}' \gets \text{update}(L, \bm{\psi}')$ \Comment{Update $\bm{\psi}'$ with Adam using gradient $\partial L / \partial \bm{\psi}'$}
            \EndFor
        \EndFor
    \end{algorithmic}
\end{algorithm}

\clearpage
\section{\label{sec:gravitational}Gravitational waves}
Posterior inference for GW events via NPE is possible for BBH events with chirp masses above 15 solar masses~\cite{dax2023neural}. However, extending SBI models to low-mass BBH events has proven challenging. Here, we will show that by adapting an effective prior and by fine-tuning the NPE model for given events, it becomes possible to accurately infer posteriors for BBH events with chirp masses between 5 and 100 solar masses. This section is structured as follows: first, we discuss the choice of prior in gravitational wave inference. Second, the data generation and preprocessing steps are discussed. Third, we discuss the incorporation of symmetry relations into the fine-tuning procedure to ensure all modes of the posterior are found. Finally, we present and discuss the inference results for simulated, non-precessing BBH GW events with a chirp mass between 5 and 100 solar masses.

\subsection{Effective priors for gravitational waves}
\begin{figure}
    \centering
    \includegraphics[width=0.49\linewidth]{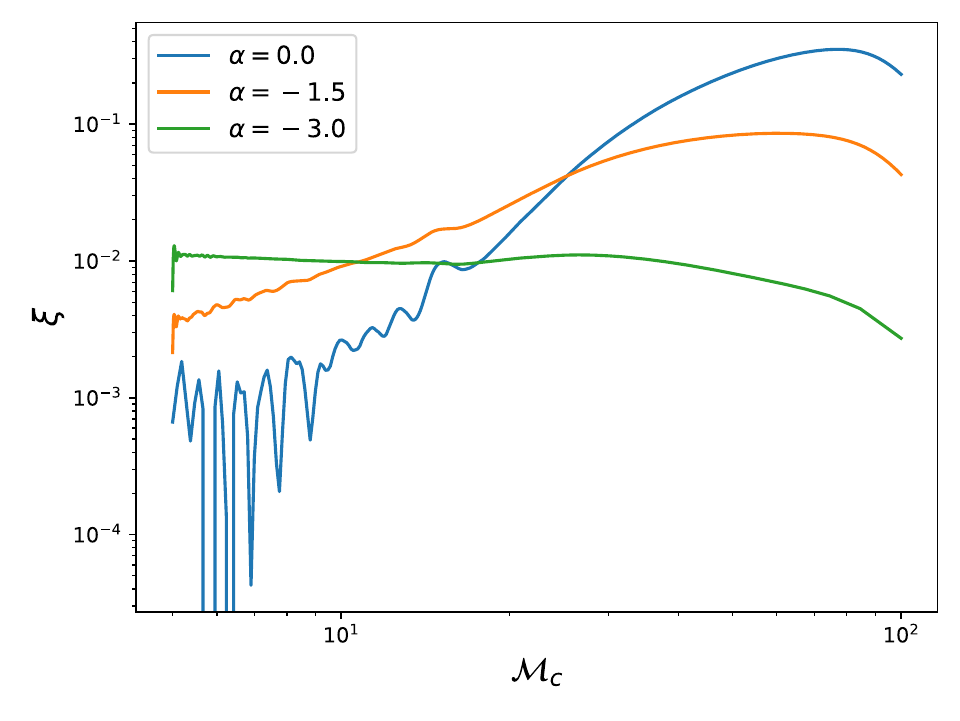}
    \hfill
    \includegraphics[width=0.49\linewidth]{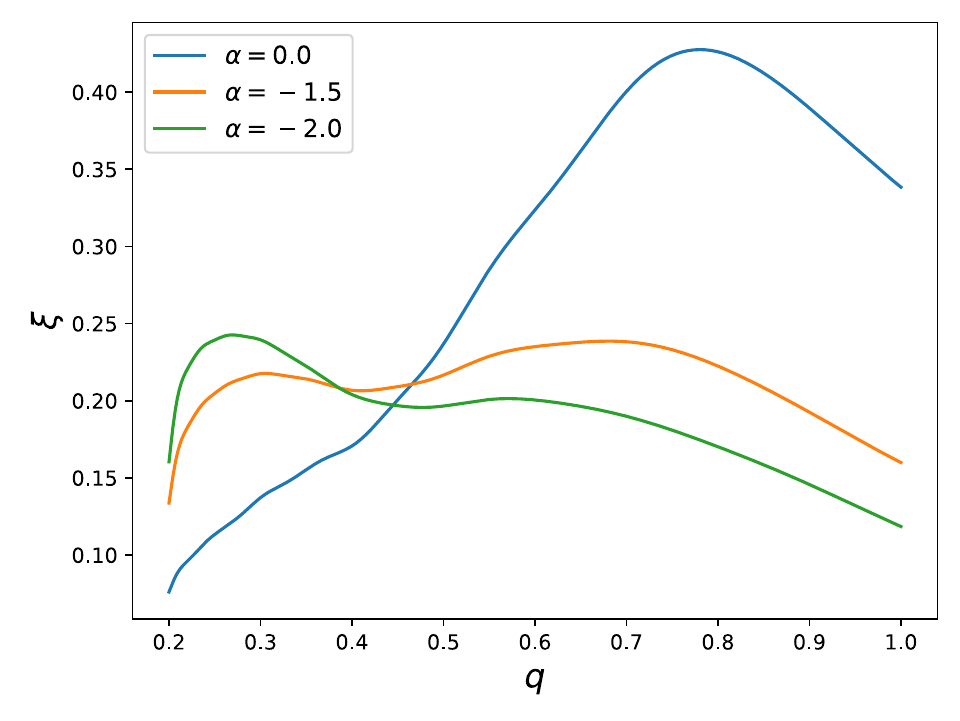}
    \caption{The mean sample exposure as a function of the chirp mass $\mathcal{M}_c$ and mass ratio $q$. Switching from a uniform prior to a power law distribution improves sample exposure for both the chirp mass and the mass ratio.}
    \label{fig:gw_eff_prior}
\end{figure}

One of the most commonly chosen priors for BBH events has uniform distributions for the chirp mass and mass ratio. Previous works in machine learning for gravitational wave inference commonly adopt either uniform priors for chirp mass and mass ratio or uniform priors for the component masses~\cite{dax2021real, bhardwaj2023peregrine, kolmus2022fast, langendorff2023normalizing}. As shown in section~\ref{sec:effective_priors}, NPE model performance matches the sample exposure caused by the choice of prior. In the left graph of Figure~\ref{fig:gw_eff_prior}, we show the sample exposure as a function of chirp mass. By switching from a uniform prior to a power law with $\alpha = -3$, the sample exposure is evenly distributed across the chirp mass range. To put the difference in GW similarity into context: the average match between two gravitational waves with chirp masses of 5.000 and 5.025 is about equal to the average match of two gravitational waves with chirp masses of 60 and 90. A similar analysis can be performed for the mass ratio; the results are shown in the right graph of Figure~\ref{fig:gw_eff_prior}. While a power law as prior may not result in a uniform sample exposure, significant improvement is achieved by choosing a power law with $\alpha=-1.5$. To improve the sample diversity during training, we chose a power law with $\alpha = -3$ for the chirp mass and $\alpha = -1.5$ for the mass ratio.

\subsection{Data generation}
The full prior from which parameter samples were generated is specified in Table~\ref{table:data_prior_gw}. For the generation of the waveforms we used the \textsc{IMRPhenomXAS}\ waveform model~\cite{pratten2020setting} provided by the \textsc{ripple} library~\cite{edwards2023ripple}, which allowed us to run the waveform generation on a GPU. The training waveforms were generated in the frequency domain between 20 and 256 Hz and with a duration of 24 seconds. The selected frequency range was chosen to optimize data generation and reduce memory burden during training. This range, while not covering the entire potential frequency range of low-mass BBH mergers, is deemed suitable for training the NPE model to capture a rough posterior. During the fine-tuning procedure, we generate waveforms in the 20 to 2048 Hz frequency band to ensure that the model converges to the correct posterior. To speed up data generation further, we use each generated waveform eight times, with each use featuring a new sky position, signal-to-noise ratio (SNR), and polarization angle. To ensure that detectable signals are given to the NPE model, we sample the optimal SNR from a uniform distribution between 10 and 30 and scale the luminosity distance to match the sampled SNR. All waveforms were whitened with the design sensitivity power spectral densities of the HLV detectors~\cite{collaboration2015advanced, acernese2014advanced, akutsu2021overview}. These steps allowed us to quickly and continuously generate parameter-strain pairs during training to prevent overfitting.

\subsection{Reduced-order basis}
The NPE model trained on the generated data remains the same as specified in section~\ref{sub:model}. However, its input is not the raw frequency series, but the frequency series projected on a reduced-order basis (ROB). While building an ROB for GWs is regularly performed with singular value decomposition (SVD)~\cite{cannon2010singular}, our approach utilizes the covariance matrix and its eigendecomposition. This choice was made to accommodate the large number of samples that are required to guarantee strong coverage. Computing the covariance matrix and its eigendecomposition consumes constant memory with respect to the number of samples. Consequently, the ROB can be constructed with as many samples as necessary to achieve sufficient coverage. Our ROB was made by calculating, per detector, the eigenbasis of the covariance matrix over five million simulated strains and taking the first 768 eigenvectors. This ROB had a minimal match of 0.95 when tested on a million samples. The ROB increases the convergence of the NPE model by providing rudimentary denoising of the frequency series. 

\subsection{Fine-tuning for gravitational wave inference}
The fine-tuning procedure is a slightly augmented version of Algorithm~\ref{algo:fine_tuning} --- we redefine our $R(\bm{\theta}|\bm{\psi}', \bm{c})$ to incorporate the potential symmetries in the polarization-phase ($\psi$-$\phi_c$) plane. For each sample drawn from $R(\bm{\theta}|\bm{\psi}', \bm{c})$, three additional copies are introduced, each shifted by $\pi$ in $\phi_c$ and/or $0.5\pi$ in $\psi$ to encompass all four potential modalities. To reflect this symmetry in $R$, we average the likelihood of the four samples, assigning this average likelihood to all four instances. This approach safeguards against missing modes due to unfortunate sampling or inaccuracies in predictions from the amortized NPE model.

As already shown in \ref{app:fine_tuning_conditioning} adding additional flow layers before fine-tuning improves the performance of the model. From our experience, the effect is not as pronounced for GWs, however it is still a positive effect. 

We fine-tune the NPE model for 20 cycles. In each of the first 10 cycles, we generate 100000 strains with a frequency range spanning from 20 to 256 Hz. In these cycles, the initial rough posterior concentrates its probability mass in the correct parts of the parameter space but does not necessarily match the true posterior perfectly. In each of the remaining cycles, we generate 50000 strains with a frequency range spanning from 20 to 2048 Hz. In this second phase, the likelihood contributions of the higher frequencies correct the posterior prediction where needed. Afterward, the NPE model generates samples, which are importance-weighted, until 100000 samples are generated or the effective sample size reaches 5000. The entire fine-tuning procedure takes 10 minutes at most on an NVIDIA GeForce RTX 3090, including model loading, \textsc{JAX} compilation, and the importance sampling after fine-tuning. 

\begin{figure}
    \centering
    \includegraphics[width=\linewidth]{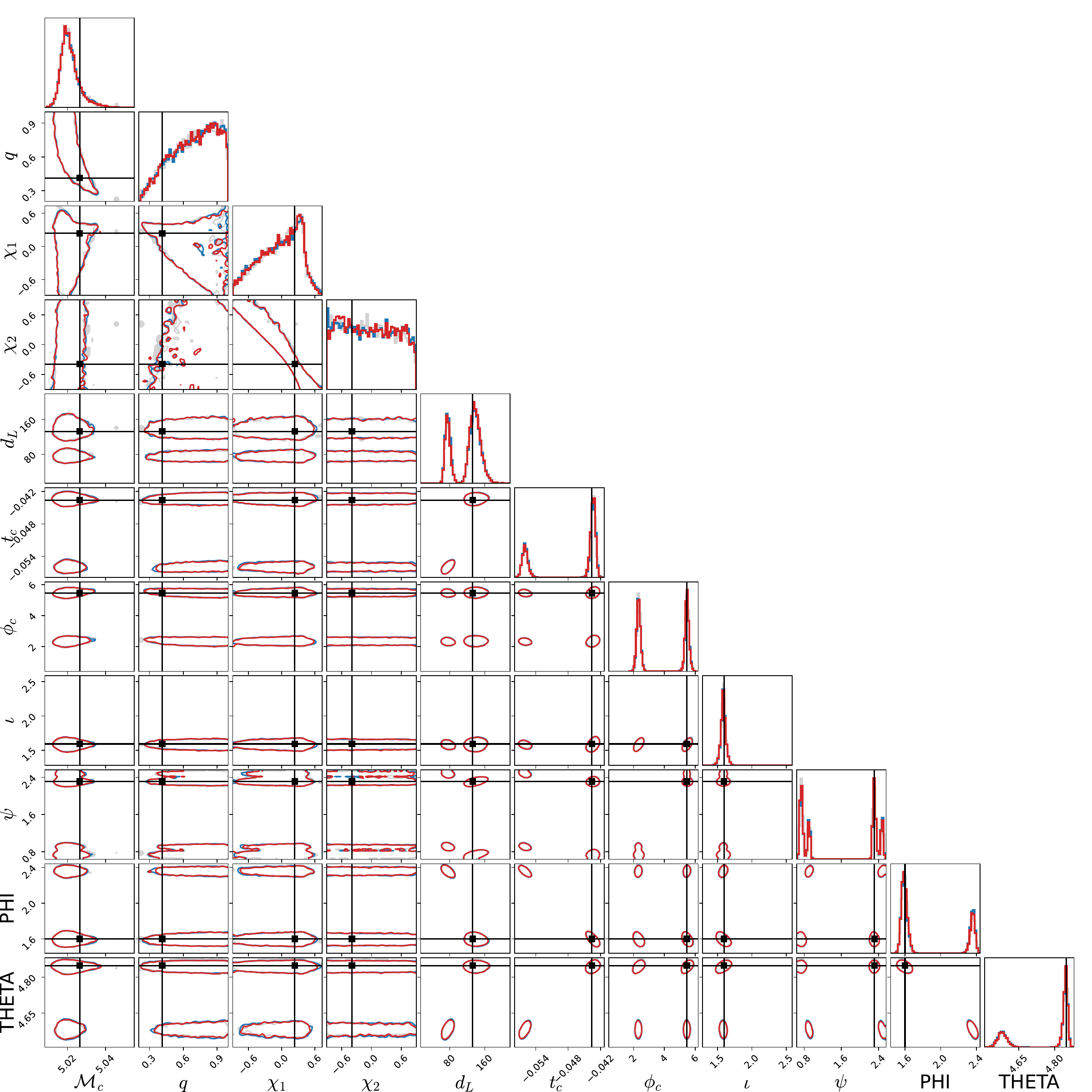}
    \caption{The predicted posterior distribution for a low-mass O3 signal, shown as 1D histograms and 2D contour plots. The contours in blue depict the predictions of the fine-tuned NPE model and the contours in red the importance-sampled posterior distribution. The grey mass is the posterior distribution obtained via nested sampling, for easy comparison we choose to use a filled contour. The NPE posterior matches the nested sampling posterior quite well. Moreover, despite the many modes in the posterior distribution, the fine-tuning procedure is still able to find all of them. The sample efficiency of amortized NPE and after fine-tuning NPE for this event differs by a factor of 20000 ($0.0025\%$ vs $51.2\%$). }
    \label{fig:gw_diff_post}
\end{figure}

The fine-tuned posteriors often closely resemble the importance-sampled ones. In Figure~\ref{fig:gw_diff_post}, we can see that the fine-tuned posterior (blue) closely aligns with the importance-sampled posterior (red). Despite the challenging characteristics of this event—featuring a low chirp mass, mass ratio, and high multimodality—the fine-tuned NPE model accurately captures the posterior distribution. Notably, just 10 minutes of fine-tuning results in a significant increase in sample efficiency, increasing from $0.00249\%$ to $51.2\%$. We see similar performance across the entire parameter space. An example of a low-sample efficiency posterior is shown in \ref{app:extra_post}. To compare with Bayesian inference methods a posterior distribution was inferred with nested sampling. For a fair comparison, the nested sampling algorithm was implemented in \textsc{JAX} to have access to GPU waveform generation. The implementation is based on RADFRIENDS~\cite{buchner2016statistical}, due to the ease of implementation and robustness of its results. The nested sampling posterior, shown as a filled grey contour in Figure~\ref{fig:gw_diff_post}, closely aligns with the importance sampled posterior. However, the time to compute the posterior distribution with nested sampling is more than 3 days, 400 times longer than our fine-tuning algorithm. To be fair, RADFRIENDS is not the most time-efficient nested sampling implementation and we expect more sophisticated implementations can complete the posterior inference within a day. 

\begin{figure}
    \centering
    \includegraphics[width=0.6\linewidth]{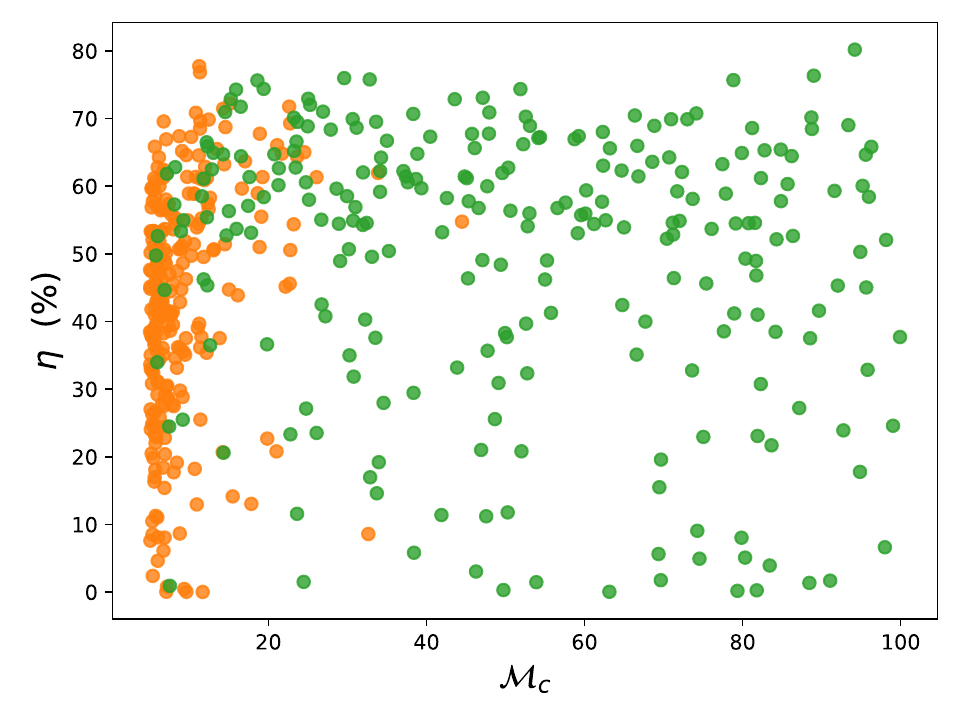}
    \caption{The sample efficiency after fine-tuning and chirp mass over 500 simulated GW events. Half of these events were sampled from a power law (orange), and the other 250 events were sampled from a uniform distribution (green). Across the entire chirp mass range we achieve strong performance. The near-zero sampling efficiencies are due to large importance weights that dominate the sampling efficiency or not-yet fully converged posteriors, for examples see~\ref{app:extra_post}.}
    \label{fig:gw_summary}
\end{figure}

To quantify the NPE model performance after fine-tuning, we simulated 500 GW events and predicted their posterior distributions by fine-tuning the NPE model for the events. The corresponding sampling efficiencies and chirp masses are depicted in Figure~\ref{fig:gw_summary}. For half of the 500 GW events, the chirp mass was drawn from a power-law distribution (shown in orange), while for the remaining 250 events, the chirp mass was drawn from a uniform distribution (depicted in green). Across the 500 events, 14 events exhibited a sample efficiency below 5\%. For 11 out of the 14 events, the low sample efficiency can be attributed to the NPE model assigning a low likelihood to a high likelihood sample, reducing the sampling efficiency. The loss of the remaining 3 events did not converge within the two rounds and required an additional ten cycles for convergence. These events can easily be identified by a high percentage of near-zero importance weights. 

\clearpage
\section{\label{sec:conclusion}Conclusion}
In conclusion, our investigation into GW inference using NPE models has yielded promising results that advance the capabilities of SBI methods. The performance of NPE models appears to align closely with the sample exposure, stressing the importance of prior selection. Moreover, our fine-tuning approach proves pivotal in overcoming the inherent limitations of amortized NPE models, providing a pathway to accurate inference for low-mass BBH posteriors. Although our primary focus is on GW inference, we believe that our findings may prove fruitful in other research areas. 

Looking ahead, we see many avenues for further improvement in fine-tuning for GW events. The implementation of adaptive stopping mechanisms holds promise to enhance convergence speed, allowing us to monitor loss or adjust frequency ranges based on initial chirp mass estimates. Differentiable waveforms enable us to use score matching to reduce the number of cycles required during fine-tuning, saving even more valuable time.  Additionally, the likelihood can be chosen at the start of fine-tuning, removing problems of unseen power spectral densities, or changing different waveform models after convergence. As part of future work, we also aim to explore even longer signal durations and to go to even lower chirp masses by considering NSBH or BNS events. 

\ack
We are thankful to Gr\'{e}gory Baltus and Iordan Ganev for insightful discussions and support. We also thank Michael Williams for the careful re-reading of the manuscript. We acknowledge the Python community~\cite{van1995python} and the core set of tools that enabled this work, including numpy~\cite{harris2020array}, \textsc{Jax}~\cite{jax2018github}, Equinox~\cite{kidger2021equinox}, \textsc{ripple}~\cite{edwards2023ripple}, bilby~\cite{ashton2019bilby}, matplotlib~\cite{Hunter:2007}, and corner~\cite{corner}. A.K. is supported by the NWO under the CORTEX project (NWA.1160.18.316). J.J. and T.B are supported by the research program of the Netherlands Organisation for Scientific Research (NWO).

\clearpage
\bibliographystyle{unsrt} 
\bibliography{main}

\clearpage
\appendix

\section{\label{app:experimental_details}Experimental details}

\subsection{\label{app:sub:oscillator}Coupled-harmonic oscillators}
The context network consisted of a linear layer, three residual blocks, and a linear layer. The initial linear layer reduced the dimension from $\text{number of oscillators} \times \text{duration} \times \text{sampling frequency}$ down to 512. The residual blocks contained an MLP following the pre-activation format suggested in reference~\cite{he2016identity}. Specifically, the MLP was defined by the following sequence a GELU activation function~\cite{hendrycks2016gaussian}, a LayerNorm~\cite{ba2016layer}, a linear layer with an output dimension of 512, followed by a GELU activation, a LayerNorm, and linear layer with an output dimension of 512. The final linear layer reduced the dimension from 512 to 128. The weight vectors of the linear layers were reparamertized following the weight normalization paper~\cite{salimans2016weight}, which significantly improved convergence rates. 

Each coupling layer in the NF model had its own MLP to parameterize its 128-degree Bernstein polynomial. These MLPs consisted of a linear layer with an output dimension of 256, a GELU activation function, a LayerNorm, and another linear layer with an output dimension of 128. It was observed that the use of weight normalization did not improve convergence in this context, so it was not used.

For each training step, we generated a new data batch, by sampling the parameters $\bm{\theta}$ from the prior specified in Table~\ref{table:data_prior_oscillator} and running the simulations to generate the corresponding oscillations and finally adding white noise. Each batch consisted of 1024 parameters-observation pairs and each epoch had 5000 training steps. The NPE model was optimized using Adam~\cite{kingma2014adam}, with a learning rate of 0.01 for the first 90\% of the epochs, and 0.001 for the last 10\% of the epochs. Lowering the learning rate further gave minimal improvements so the cut-off was set at a learning rate of 0.001.

\begin{table}[ht]
    \centering
    \caption{The priors used for the data generation of the coupled-harmonic oscillators. The prior for the mass is a power law prior whose coefficient was either -3.0, -1.5, or 0.0, depending on the experiment.}
    \label{table:data_prior_oscillator}
        \begin{tabular}{lcccc}
        parameter                   & prior                         & minimum               & maximum   & unit \\ \hline
        $m$                        & Power law (-3.0, -1.5, 0.0)    & 0.1                    & 10       & kg   \\ 
        $k$                        & Uniform                        & 10                     & 100      & N/m   \\ 
        $a_{0, 1, 2, 3}$           & Uniform                        & 0.5                    & 5.0      & m  \\ 
        $\phi_{0, 1, 2, 3}$        & Uniform                        & 0                      & 2$\pi$   & rad   
        \end{tabular}
\end{table}

\clearpage
\subsection{\label{app:sub:gw}Gravitational waves}
The context network is identical in setup compared to the one made for the coupled-harmonic oscillator, except that the dimensions are bigger. The all linear layer, except the last linear layer, has an output dimension of 4096. The last linear layer has an output dimension of 512. The setup of the NF model is the same for the coupled-harmonic oscillator. 

For each training step, we generated a new data batch, by sampling the parameters $\bm{\theta}$ from the prior specified in Table~\ref{table:data_prior_gw}. To improve the generation speed we generate only 64 waveforms, and each waveform is copied 16 times and gets new sky coordinates, polarization angle, SNR, and arrival time which are used to scale, and subsequently, project the waveform onto the HLV detectors. The waveforms are then whitened after which we add white noise.

Each batch consisted of 1024 parameters-observation pairs and each epoch was 5000 training steps. The NPE model was optimized using Adam~\cite{kingma2014adam}, with a learning rate of 0.01 for the first 450 epochs, and 0.001 for the last 50 epochs. Lowering the learning rate further gave minimal improvements so the cut-off was set at a learning rate of 0.001.

For the sky coordinates we choose to use the polar coordinates over the celestial coordinates removing the implicit dependence on Greenwich Mean Sidereal Time. Since we already have several theta's and phi's as notation we opted to use THETA and PHI as notation for these polar / sky coordinates.

\begin{table}[ht]
    \centering
    \caption{The priors used for the data generation of the GWs. Instead of luminosity distance the optimal SNR of the signal is sampled. During the generation of the waveform the luminosity distance is set to 1000 MPC and after the generation, the waveform and luminosity distance are scaled to match the desired SNR.}
    \label{table:data_prior_gw}
        \begin{tabular}{lccccc}
        Parameter                   & Name                  & Prior             & Minimum               & Maximum   & Unit \\ \hline
        $m_1, m_2$                  & Component mass        & Constraint        & 3                     & 150       & $\textup{M}_\odot$ \\ 
        $\mathcal{M}_c$             & Chirp mass            & Power law (-3.0)  & 5                     & 100       & $\textup{M}_\odot$ \\
        $q$                         & Mass ratio            &  Power law (-1.5)  & 0.2                   & 1.0       & -    \\
        $|\chi_1|,\, |\chi_2|$      & Spin amplitudes       & Uniform           & 0                     & 0.9       & -    \\ 
        THETA                       & Sky coordinate 1      & Uniform           & 0                     & 2$\pi$     & rad  \\ 
        PHI                         & Sky coordinate 2      & Cosine            & 0                     & $\pi$    & rad  \\ 
        $t_c$                       & Coalescence time      & Uniform           & -0.1                  & 0.1       & s    \\ 
        $\phi_c$                    & Coalescence phase     & Uniform           & 0                     & 2$\pi$    & rad  \\ 
        $\iota$                     & Inclination angle     & Sine              & 0                     & $\pi$     & rad  \\ 
        $\psi$                      & Polarization angle    & Uniform           & 0                     & $\pi$    & rad  \\ 
        SNR                         & Signal-to noise ratio & Uniform           & 10                    & 30        & -
        \end{tabular}
\end{table}

\clearpage
\section{\label{app:fine_tuning_conditioning}Neural conditioning and fine-tuning}
There are several straightforward ways one can condition the coupling layers on the output of the context network. Perhaps the easiest is to concatenate the context vector and the static half $\bm{b}^l_j$ and feed the new vector to the MLP of the coupling layer. As we can see in Table~\ref{tab:conditioning_perfomance}, it works well for training, but it is not optimal for fine-tuning. Slightly more involved methods transform the context vector, via a linear transformation, into a bias vector, a scaling vector followed potentially by a sigmoid function. Of these three conditioning methods, a scaling vector is the most optimal solution when it comes to fine-tuning performance. Throughout this paper, the scale method of conditioning was used.

\begin{table}[h]
    \centering
    \caption{The performance of NPE models with different forms of conditioning. They were trained for 5 epochs on the toy problem, with a duration of 2 seconds. }
    \label{tab:conditioning_perfomance}
    \begin{tabular}{lccc}
    Conditioning & Training loss & Before fine-tuning (\%) & After fine-tuning (\%) \\ \hline
    Concatenate  &   -11.5       &  0.76 &  39.1  \\
    Bias         &   -11.7       &  0.76 &  38.9  \\
    Scale        &   -11.2       &  0.60 &  47.5  \\
    Sigmoid      &   -11.5       &  0.76 &  41.2                                   
    \end{tabular}
\end{table}

Fine-tuning allows us to add NF layers after the amortized training. These do not need conditioning and can improve the flexibility of the NPE model. Since the NF layers are initialized such that they approximate the identity function they should not alter the output too much. In Table~\ref{tab:additional_layer} we report the improvements in fine-tuning an NPE model with a variable number of coupling layers. The NPE model was trained for 100 epochs on coupled-harmonic oscillators with signals of 20 seconds. 

\begin{table}[h]
    \centering
    \caption{The results of fine-tuning for with additional layers, }
    \label{tab:additional_layer}
    \begin{tabular}{cc}
    Additional coupling layer & After fine-tuning (\%) \\ \hline
    0  &  2.1  \\
    1  &  4.4  \\
    2  &  7.9  \\
    4  &  8.3  \\
    8  &  7.9  \\
    \end{tabular}
\end{table}

\clearpage
\section{\label{app:extra_post}Outlier in GW posterior}
\begin{figure}[h]
    \centering
    \includegraphics[width=\linewidth]{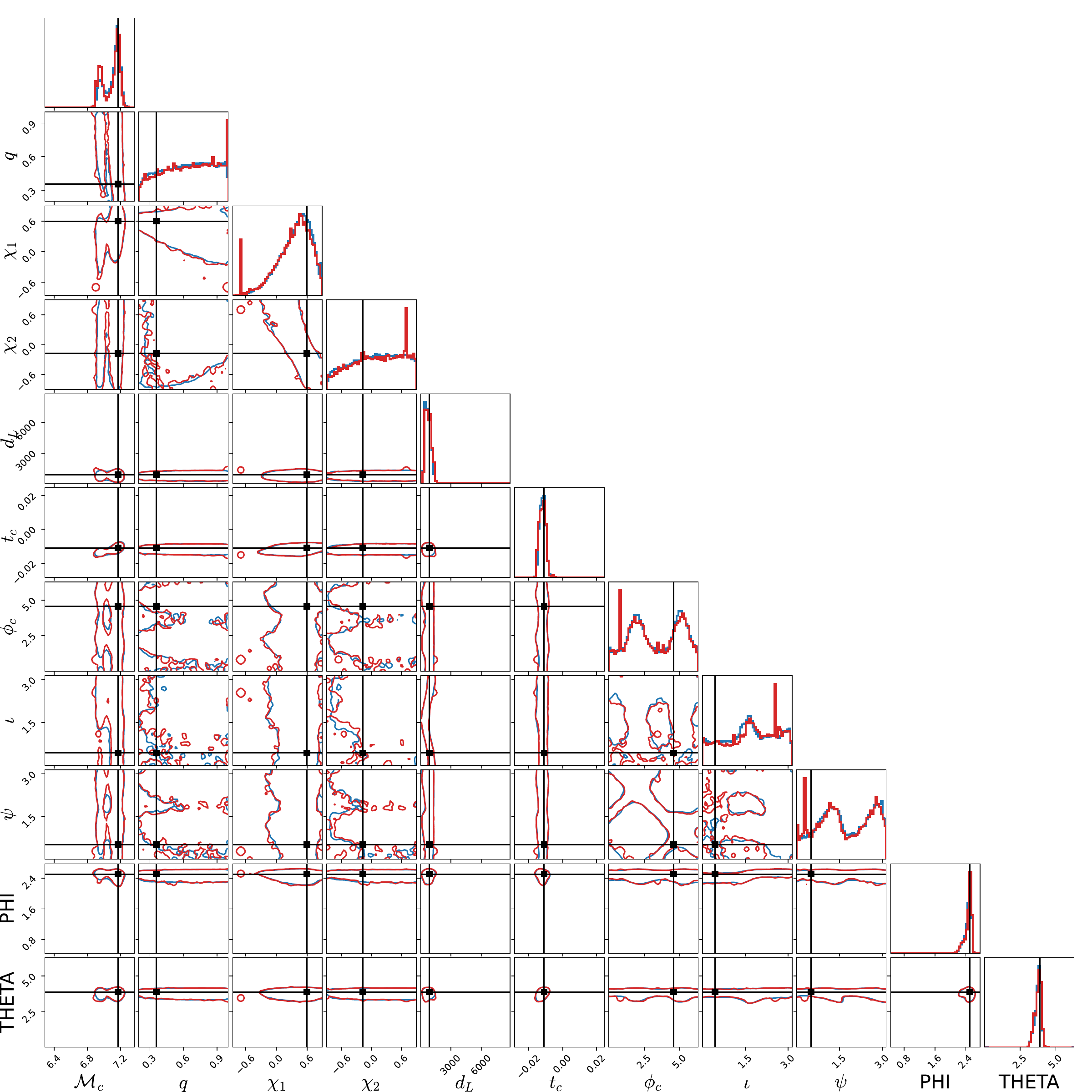}
    \caption{A GW posterior with a sample efficiency of 0.79\%. Although it seems to have found the posterior distribution, for a single sample the ratio between actual and assigned likelihood is massive, resulting in a low sample efficiency. This sample shows up in the 1D histograms as a sharp peak in red.}
    \label{fig:gw_outlier}
\end{figure}

In Figure~\ref{fig:gw_outlier} a typical low sample efficiency posterior is shown. Although the NPE model seems to capture the posterior quite well, there is a single sample that has a very large weight which lowers the sample efficiency. Another fine-tuning cycle can correct the discrepancy.

\end{document}